# A CHALLENGE TO CONTROL GRAVITY VIA APPLYING ELECTROMAGNETIC LOW-FREQUENCY RADIATION – THEORY AND PROPOSED MODEL EXPERIMENTS

Július Vanko[a], Miroslav Súkeník[b] and Jozef Šima[b]

[a]Comenius University, FMPI, Mlynská dolina, Bratislava, [b]Slovak Technical University, FCHPT, Radlinského 9, 812 37 Bratislava, Slovakia

vanko@fmph.uniba.sk; sukenik.miroslav@stonline.sk;

Abstract

Including Vaidya metric into the model of Expansive Nondecelerative Universe allows to localize the energy of gravitational field. A term of effective gravitational range is introduced and classic Newton potential is substituted for Yukawa-type potential. It allows to allocate a typical frequency value to each gravitational field. Derived theoretical conclusions led us to investigate the effect of electromagnetic field with a precisely predetermined frequency and intensity on iron. We believe that under certain circumstances a decrease in iron gravitational mass should be observed. Two model experiments verifying the theoretical conclusions are proposed.

1. Introduction

Among the main parameters describing the fundamental physical field, its static potential $\phi$ should be introduced. This potential contains a dumping function $e^{-\delta r}$, in which $\delta$ means a reciprocal value of the Compton wavelength of the corresponding boson mediating a given interaction. The dumping function plays a fundamental role for the short-range fields. For both the gravitational and electromagnetic interactions the value of the function is close to 1. Having, however, a medium with a higher amount of charged particles, the dumping function for electromagnetic interaction adopts the form $e^{-r/d}$, where d is so-called Debye length. It means that the potential of a given particle is shielded by its neighbouring charged particles and in a distance r above d it approaches zero value. All these facts are well-known.

In this contribution a new dumping function, namely $e^{-r/r_{ef}}$, is introduced for gravitational field, where $r_{ef}$ expresses the effective gravitational range of a body. Its meaning lies in the presumption that in the distance r exceeding the effective range, the gravitational potential of a body is shielded by the neighbouring mass and drops down to non-measurable values. Provided that the mean Universe mass density is known (and it is known relatively precisely), this effective range may be determined for any body. Yukawa potential is thus applicable for all fundamental physical interactions. It is obvious that in spite of the fact that the dumping function may play an important role for the long-range interactions, the corresponding mediating bosons (gravitons and photons) are of zero rest mass. To put the effective gravitational range and Compton length of graviton identical is thus only an analogy serving to better understand the issue.

2. Model of Expansive Nondecelerative Universe

Starting from the beginning of 80's, the inflationary model of the universe acquired dominant position in cosmology [1]. The model has been able to eliminate certain cosmological problems, at the same time it has, however, open new questions, such as the Universe age, Hubble's constant or deceleration parameter values. Its weak point lies in the fact that it has not contributed to deepen our understanding of the gravitation and its relation to the other physical interactions. Moreover, in accordance with some analyses [2], the initial nonhomogenities should not be eliminated but they are rather enhanced within the period of the Universe inflation.

Open questions have been a challenge for developing further models of the Universe, one of them being the Expansive Nondecelerative Universe (ENU) model.

The cornerstones of ENU have been discussed in detail elsewhere [3 - 9] and can be summarized as follows:

a) The Universe, throughout the whole expansive evolutionary phase, expands by a constant rate equals the speed of light c obeying thus relation

$$a = ct_U = 2Gm_U/c^2 \qquad (1)$$

where a is the Universe radius (equals to the Universe gauge factor in the ENU), $t_U$ the Universe age, $m_U$ is the Universe mass (their present ENU-based values are following: $a = 1.3 \times 10^{26}$ m, $m_U = 1 \times 10^{53}$ kg, $t_U = 1.37 \times 10^{10}$ yr).

b) The curvature index k and Einstein cosmologic constant $\Lambda$ are of zero value

$$k = 0 \qquad (2)$$

$$\Lambda = 0 \qquad (3)$$

c) The mean energy density of the Universe is identical just to its critical density.

d) Since a is increasing in time, $m_u$ must increase as well, i.e. in the ENU, the creation of matter occurs. The total mass-energy of the Universe must, however, be exactly zero. It is achieved by a simultaneous gravitational field creation, the energy of which is negative. The fundamental mass-energy conservation law is thus observed.

e) Due to the matter creation, Schwarzschild metrics must be replaced in ENU by Vaidya metrics [10 - 12] in which the line element is formulated as

$$ds^2 = \frac{d\psi^2}{c\,dt} \cdot \frac{1}{f_{(m)}^2} \left(1 - \frac{2\psi}{r}\right) c^2 dt^2 - \left(1 - \frac{2\psi}{r}\right)^{-1} dr^2 - r^2\left(d\theta^2 + \sin^2\theta\, d\phi^2\right) \tag{4}$$

and the scalar curvature R (which is, contrary to a more frequently used Schwarzschild metric of non-zero value in Vaidya approach also outside the body allowing thus to localize the gravitational energy density) in the form

$$R = \frac{6G}{c^3 r^2} \cdot \frac{dm}{dt} = \frac{3\, r_g}{a\, r^2} \tag{5}$$

where m is the mass of a body, G ($6.67259 \cdot 10^{-11}$ kg$^{-1}$ m$^3$ s$^{-2}$) is the gravitational constant, r is the distance from the body, $r_g$ is the gravitational radius of the body, $f_{(m)}$ is an arbitrary function, and $\Psi$ is defined [7] as

$$\Psi = \frac{G\,m}{c^2} \tag{6}$$

In order to $f_{(m)}$ be of nonzero value, it must hold

$$f_{(m)} = \Psi \qquad \frac{d}{dr}\left(1 - \frac{2\Psi}{r}\right) = \frac{2\Psi^2}{r^2} \tag{7}$$

Based on (1), in the framework of the ENU model

$$\frac{d\Psi}{c\,dt} = \frac{\Psi}{a} \tag{8}$$

Dynamic character of the ENU is described by Friedmann equations. Introducing dimensionless conform time, equation (1) can be expressed as

$$c\,dt = a\,d\eta \tag{9}$$

from which

$$a' = \frac{da}{d\eta} \tag{10}$$

Applying Vaidya metric and stemming from Robertson-Walker approach, Friedmann equations [13] can be then written in the form

$$\frac{d}{d\eta}\frac{1}{a}\cdot\frac{da}{d\eta} = -\frac{4\pi G}{3c^4}a^2(\varepsilon + 3p) \tag{11}$$

$$\frac{1}{a}\cdot\frac{da}{d\eta}^2 = \frac{8\pi G}{3c^4}a^2\varepsilon - k \tag{12}$$

where $\varepsilon$ is the critical energy density (identical to the actual density within the ENU model) and p is the pressure. Based on (11) and (12) it follows for energy density of the Universe within the ENU

$$\varepsilon = \frac{3c^4}{8\pi Ga^2} \tag{13}$$

$$p = -\frac{\varepsilon}{3} \tag{14}$$

Equations (13) and (14) represent the matter creation and the negative value of gravitational energy, respectively (for more details, see [3 - 5]).

A typical feature of the ENU model lies in its simplicity, in the fact that no „additional parameters" or strange „dark energy" are needed, and in the usage of only one state equation in the Universe description. Calculated gauge factor a, cosmological time $t_u$, and energy density $\varepsilon$ match well the generally accepted values.

Owing to Vaidya metric application, the ENU model enables to localize gravitational energy [7]. Stemming from a general formula [7], the absolute value of gravitational energy density $\varepsilon_g$ is a function of a body mass m and the distance r from the body

$$|\varepsilon_g| = \frac{Rc^4}{8\pi G} \cong \frac{3mc^2}{4\pi a r^2} \tag{15}$$

The absolute value of the gravitational field density of a system equals to the critical density at the distance $r_{ef}$ (effective interaction range of its gravitational force). Comparison of the relations (13) and (15) leads to

$$r_{ef} = (r_g a)^{1/2} = (2\psi a)^{1/2} \qquad (16)$$

in which $r_g$ means the gravitational radius of a body with the mass m.

Vaidya metric may be applied in all cases for which the gravitational energy is localizable, i.e. in cases being governed by relation (17)

$$r < r_{ef} \qquad (17)$$

Gravitational influence can be thus actually realized only if the absolute value of the gravitational energy density created by a body exceeds the critical energy density, i.e. the mean gravitational energy density of the Universe. If

$$r \geq r_{ef} \qquad (18)$$

Vaidya metric adopts the form of Schwarzschild metric preventing the energy of gravitational field from localization.

Taking Vaidya metric into account, the non-relativistic gravitational potential outside a body with the mass m can be expressed in the form

$$\Phi = \Phi_0 \exp(-r / r_{ef}) \qquad (19)$$

where

$$\Phi_0 = -G \int \frac{\rho}{r} dV = -\frac{G m}{r} \qquad (20)$$

3. Gravitational Yukawa potential and its consequences for micro-world and cosmos

In a certain approximation, the effective range $r_{ef}$ may be understood as Compton wavelength of the interchanged particle mediating a corresponding physical interaction (in such a case, $r_{ef}$ represents the graviton wavelength and $c/r_{ef}$ its frequency). The heavier the body, the longer length of its effect (the longer its effective range) and, in turn, the lower frequency must have the related quantum of gravitational field.

From the viewpoint of theory, there is not limit in the range of gravity, however, in a distance r exceeding the given effective range $r_{ef}$, the body gravitational effect strongly drops and becomes immeasurable.

The Universe effective range as a whole equals to its gauge factor a (about $1.3 \times 10^{26}$ m). The related frequency approaches $2 \times 10^{-18}$ Hz (being a reciprocal value of the cosmologic time, 14 billion years). There is no physical or logic meaning in the assumption that the gravitation in a visible Universe part acts in a longer distance as the distance of light passing during the cosmological time.

The effective range of our Earth and Sun is about $10^{12}$ m (frequency $2.8 \times 10^{-4}$ Hz ) and $6.2 \times 10^{14}$ m ( frequency $4.8 \times 10^{-7}$ Hz ), respectively.

A mass of the lightest particle able to exert gravitational force to its environment is about $10^{-28}$ kg. Its Compton wavelength and effective gravitational range are identical and reach at present $10^{-15}$ m (frequency $10^{23}$ Hz ). All particles with a greater mass can exert gravitational force to their surroundings, all lighter particles will do so later (see eq. 16 documenting a dependence of effective range on the gauge factor, i.e. on the cosmologic time too).

The gravitational effective range of the proton is at present about $1.8 \times 10^{-14}$ m (frequency $1.67 \times 10^{22}$ Hz ). It is very interesting that just in the beginning of the matter era the effective range of the proton was the same as its Compton wavelength. It follows, it was just the time when the proton began to exert its force on its surroundings. This was of primary importance and merit at the formation of first gravitational fluctuations and nonhomogenities.

For the majority of atomic nuclei and the heaviest particles which can be formed at the time being (bosons Z and W ),the effective gravitational range reaches about $10^{-13}$ m (frequency $3 \times 10^{21}$ Hz ). It means that if we want to investigate the gravitation in the micro-world, we have at our disposal only the length region spanning from $10^{-15}$ m to $10^{-13}$ m. Putting LHC in CERN into operation, the upper limit will approach $10^{-12}$ m. Any effort to investigate gravitation beyond $10^{-12}$ m is out of question at present (and the situation will not change in near future until creation of more effective accelerators).

At the end of this part, one interesting fact is worth mentioning. The effective range of the neutron is currently close to $1.8 \times 10^{-14}$ m. To initiate a creation of a neutron star, the neutrons must be packed closer, i.e. the mass density must be higher than $6.9 \times 10^{10}$ g/cm$^3$. And it is just the case. It has been calculated and rationalized very precisely that for starting a creation of a neutron star, the density must be at least $4.3 \times 10^{11}$ g/cm$^3$. This density is associated with the effective range of the neutrons at about 4 billion years following the beginning of the Universe. The first neutron star of our Universe might be created approximately 10 billion years ago.

4.   Case study – experiment 1

The case study deals with the effect of low-frequency electromagnetic field on iron materials.

The characteristics of electromagnetic field, in particular its frequency f, power P and intensity I are dependent on the iron conductivity σ, permeability ∝, mass m and thickness. It holds for the metal skin layer δ:

$$\delta = (\pi f \propto \sigma)^{-1/2} \quad (21)$$

The following condition (22) is required to be met

$$r_{ef} = \delta \quad (22)$$

where $r_{ef}$ is the gravitational range of the metal sample (see eq. 16 ), and δ is its thickness.

Frequency of electromagnetic radiation to be used must comply with relation (23)

$$f_{tot} = (r^2_{ef} \pi \propto \sigma)^{-1} = (\delta^2 \pi \propto \sigma)^{-1} = c^2 / 2\, G\, m_{tot}\, a\, \pi \propto \sigma \quad (23)$$

in order the whole mass of the metal sample $m_{tot}$ be eliminated at the condition (22). Such parameters are, however, reached and maintained only with difficulty. This is why the condition (22) may be moderated as follows

$$\delta^2 = r^2_{ef} = 2\, G\, m_x\, a / c^2 \quad (24)$$

$$m_x = \delta^2 c^2 / 2\, G\, a \quad (25)$$

where $m_x$ is part of the sample to be compensate using electromagnetic radiation with the frequency corresponding to the skin layer δ.

Integration of (15) over volume and subsequent derivation of the integration product according to time lead to relation specifying the gravitational power. Electromagnetic power must be identical. It must therefore hold

$$P = m_x c^3 / a \quad (26)$$

If S is the body surface affected by electromagnetic radiation, the radiation intensity is

$$I = P / S \tag{27}$$

Provided that our sample will be exposed to electromagnetic radiation with the corresponding frequency and intensity, the body mass should be reduced by $m_x$.

An alternative approach lies in the possibility to encapsulate the electromagnetic radiation antenna directly into a body with the wall thickness $2\delta$. Frequency, power and intensity of electromagnetic radiation are obtained by similar procedure.

5. Case study - experiment 2

Let us have an iron cubic sample ($r = \delta = 5 \times 10^{-2}$ m) with the following parameters: density $\rho = 7.874 \times 10^3$ kg/m$^3$, permeability $\propto = 5000 \propto_0$, and conductivity $\sigma = 1.03 \times 10^7$ S/m. The cube mass is then $m = 0.98425$ kg.

Based on (21), the frequency of electromagnetic radiation is calculated as $f = 1.978 \times 10^{-3}$ Hz. Relation (25) leads to $m_x = 0.0129$ kg.

The power and intensity from (26) and (27) are $P = 2.7 \times 10^{-3}$ W and $I = 1.077$ J/m$^2$s, respectively. At these parameters, the cube mass should be reduced by $m_x$.

The mass reduction efficiency is $\eta = (m_x/ m) \times 100 = 1.32$ %.

Having another cubic sample from the same material with $r = \delta = 0.1$ m, the cube mass will be $m = 7.874$ kg. Applying radiation with the frequency $f = 4.9 \times 10^{-4}$ Hz, power $P = 10^{-2}$ W and intensity $I = 1.077$ J / m$^2$s, the mass reduction is $m_x = 0.052$ kg, i.e. $\eta = 0.66$ %.

With the samples made from identical material, the intensity of electromagnetic radiation does not depend on the sample dimension. The greater the cube, the higher its mass reduction in absolute value (and lower its efficiency), the higher must be electromagnetic power but lower its frequency The frequency can be reduced if a material with a higher permeability is used. It would be very stimulating to experimentally find whether increasing the electromagnetic radiation intensity will cause an increase $m_x$, i.e. the process efficiency.

6. Conclusion

It should be pointed out that in the issue of mass reduction, the main role is played by the intensity and power of electromagnetic radiation. Within the same time, the gravitational and electromagnetic energies passed through the body must be identical.

Due to the different rate of electromagnetic energy transfer through the body, the electromagnetic and gravitational frequencies are not and cannot be, however identical.

The precisely predefined electromagnetic frequency ensures only the total absorption of the electromagnetic radiation in the body and the interference of both fields. This interference should result in the body mass reduction. It is worth mentioning that there is no violation of the energy conservation law, since immediately after the termination of electromagnetic field action, the original mass is returned.

Owing to its precisely defined and known parameters, iron is ideal testing material.

It is not, however, the only way to change the mass of materials.

Stemming from the gravitational power of our Earth ($P = 1.246 \times 10^{24}$ J/s) and its surface ($S = 5.11 \times 10^{14}$ m$^2$), the intensity of gravitational radiation $I = 2.43 \times 10^9$ J/m$^2$ s can be calculated.

One might suppose that creating the equal intensity of electromagnetic radiation intensity at the Earth surface (which is no problem) a slight reduction in the mass of the bodies exposing to such a electromagnetic field should be measurable due to the interference.

This is the case of Podkletnov's phenomenon [14]. Podkletnov used superconductive emitor (YBaCuO$_x$), voltage from 0.5 MV to 2 MV, current up to $10^4$ A, pulses lasting $10^{-5}$ to $10^{-4}$ s. The generated charge reached about 0.1 C, i.e. the energy of emission represented about $10^5$ J. Taking the mentioned time into account, the power is about $2 \times 10^9$ W.

With regard to the Podkletnov's experiment arrangement (distance between the electrodes about 1m), the surface through which the discharge was transferred approached 1m$^2$.

Within the experiment the electromagnetic radiation intensity was about $2 \times 10^9$ J/m$^2$s, which is a value close to the gravitational field intensity of the Earth. We believe that the interference of the both field caused gravitational pulses having the same translation direction as the original electromagnetic pulse. Podkletnov succeded in measuring not very substantial but verifiable reduction in the mass of bodies exposed to the pulse.

It might be a case that an analogous mechanism applies in the solar corona heating. The average value of solar magnetic field is about $B = 10^{-4}$ T, in the spots, however, the value may reach up to $10^{-1}$ T. For the Sun, the calculated intensity of its gravitational radiation is $I = 6.78 \times 10^{10}$ J/m$^2$s, which corresponds to $B = 2.37 \times 10^{-2}$ T. At this intensity of magnetic field, the Sun gravitational and electromagnetic fields may interfere. Created gravitational pulses could move particles forming solar corona and heat them up to $10^6$ K. (Short time lasting mass reduction of elementary particle in the chromosphere and corona means an increase in their plasma frequency).

We would like to put stress on the necessity of performing further experiments accompanying the theoretical research. It is obvious that for building-up the quantum theory of gravity, a wider and more complex basis must be formed than the general theory of relativity.